\documentclass[twocolumn,amssymb,floatfix,aps,prb,showpacs]{revtex4}

\usepackage{epsfig} 
\usepackage{amsmath}
\usepackage{tikz}
\usepackage{float}
\usepackage{graphicx}
\usepackage{bbm}
\usepackage{times}
\usepackage{physics}

\begin{document} 

\title{Complete Solution of the Tight Binding Model on a Cayley Tree:  Strongly Localised versus Extended States}

\author{Deepak Aryal} 

\author{Stefan Kettemann} 

\address{Jacobs University, School of Engineering and Science,
  Campus Ring 1, 28759 Bremen, Germany}
\address{ Division of Advanced
  Materials Science Pohang University of Science and Technology
  (POSTECH) San 31, Hyoja-dong, Nam-gu, Pohang 790-784, South Korea}

\date{\today}

\begin{abstract} 
 The complete set of Eigenstates and Eigenvalues of the nearest neighbour tight binding model on a Cayley tree with branching number $b=2$ and $M$ branching generations with open boundary conditions is derived. We find that of the  $N= 1 +3  (2^M-1)$ total states
 only  $3 M +1$ states are extended throughout the Cayley tree.
 The remaining $N-(3 M+1)$ states are found to be strongly localised states with finite amplitudes on only a subset of sites. In particular, there are,
 for $M>1$, $3 \times 2^{M-2}$ surface 
    states which are  each antisymmetric combinations of only two sites on the surface of the Cayley tree and have energy eactly at $E=0$, the middle of the band. 
    The ground state and the first two excited states  of the Cayley tree are  found to be 
     extended states with amplitudes on all sites of the Cayley tree, for all $M$. 
  We  use the  results  on the complete set of Eigenstates and Eigenvalues to derive  the  total density of states and a local density of states. 
\end{abstract}


\maketitle


\section{Introduction} 

 Arthur Cayley introduced the Cayley tree graph as a graphical representation of the free group\cite{Cayley1878}. The Cayley tree is  a tree graph with $N$ nodes, branching number $b$ with degree  $k = b + 1$, except at surface edge nodes where $k= 1$. Since it is loop free, the dynamics on Cayley trees is amenable to exact solutions employing the transfer matrix method.  The local density of states 
 at the central site of a tight binding model on a Cayley tree has been derived analytically in Refs. \cite{Brinkmann1970,Chen1974,mahan,Eckstein2005,giacometti}.  The tight binding model for disordered fermions has been solved  analytically by the  transfer matrix method on a Cayley tree, revealing the Anderson delocalization transition for $b>1$\cite{Beeby1973,AbouChacra1973,Zirnbauer1986,Mirlin1994}. 
 For infinite number of lattice sites the Caylee tree is called Bethe lattice since Bethe' s approximation for the Ising model becomes  exact on this lattice\cite{Baxter1982}.
  The density of states for the Bethe lattice has been derived in Ref. 
  \cite{Derrida1993}. Other interacting models, in particular the Hubbard model have been studied on the Bethe lattice. Since in  the limit of $k \rightarrow \infty$ mean field theory for any model with interactions becomes exact, the formulation on the Bethe lattice has been used to study this limit in a controlled way\cite{Georges1996}. 
The problem of quasiparticle relaxation in an interacting electron system has been mapped on the localization problem  in Fock space and solved approximately by mapping it on a Cayley tree\cite{Altshuler1997}. 
Recently, the dynamics of coupled oscillators have been studied on a Cayley tree, as a model for the dynamics in distribution power grids \cite{Tamrakar2018}.

Inspite of this wide range of applications of the Cayley tree in physics, the Eigenstates and Energy Eigenvalues of the tight binding model have hardly been studied. In 2001, Mahan obtained the shell symmetric Eigenstates on a  Bethe lattice and derived from it the local density of states at the central site. However, the full basis of Eigenstates on a Cayley tree was not obtained there. 
We therefore  intend to fill this gap  in this paper by giving the analytical derivation for branching number $b=2$. The numerical analysis of this problem has recently been presented in Ref. \cite{Yorikawa2018}.
 \section{The Tight Binding Model on  a Cayley Tree}
\begin{figure}[h]
    \centering
    \includegraphics[width=0.35\textwidth]{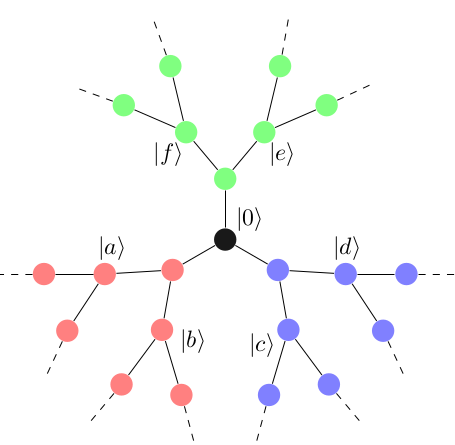}
    \caption{Cayley tree with branching $b=2$ and $M=3$ generations. Its  three branches are highlighted by different color.}
    \label{figure1}
\end{figure}

 The tight binding model
 is defined by 
 \begin{equation}\label{Eq:beast1}
   \hat{H} = \sum_{<i,j>} t_{ij} \ket{i}\bra{j}.
 \end{equation}
 where $t_{ij}$ is the hopping amplitude between sites $i$ and $j$.  $<i,j>$ denotes nearest neighbours on the  graph. Here $\ket{i}$ denotes the state in which a single particle occupies the site labeled by $i$. We will assume homogenous hopping amplitude $t_{ij} =t,$ in the following (in particular, we set $t = 1$ for simplicity). 
We  are interested in obtaining the full set of Eigenstates $\ket{\Psi_n}$ with Eigenvalues $E_n$ as given by 
 \begin{equation}\label{Eq:beast2}
   \hat{H}\ket{\Psi_n} = E_n \ket{\Psi_n},
 \end{equation}
 for all $n= 1,...,N$, where $N$ is the number of states.  Here, we consider the sites to be on a Cayley tree of branching number $b=2$, as shown in  Fig. \ref{figure1} for the example of $M=3$ branching generations, when starting from the central site. The number of sites $N$
 is related to the number of branching generations $M$. Noting that each generation $l$ has $3 \times 2^{l-1}$ sites,   $N= 1+  3 \times  \sum_{l=1}^M 2^{l-1} = 1 + 3 (2^M-1)$.

 \section{Exact Solution}
\subsection{Choice of basis}
Mahan found a subset of $M+1$ Eigenstates of all $N$ Eigenstates on a Cayley tree \cite{mahan} by using the shell symmetric states as a subbasis for the Eigenstates. These
 shell symmetric basis states are symmetric with respect to a rotation  between different branches of the Cayley tree, which are  highlighted by 
  different colors in Fig. \ref{figure1}. The Eigenstates have therefore equal amplitude on all sites of the same generation $l$.  Accordingly,
 we can  label these symmetric states with one
 number that denotes the generation $l$ of the hopping starting from the central site, e.g. $\ket{l=3}$ denotes the 
 symmetric state on all
 $3^{rd}$ generation sites. Eq. (\ref{Eq:beast2}) then furnishes recurrence relations. These were solved by Mahan to obtain solutions with equal amplitude on sites of same generation (symmetric solutions)\cite{mahan}. \par

In order to obtain all Eigenstates, we need to extend the basis to  all states to be able to  distinguish between the 
different branches of the Cayley tree. In a first step, let us 
 split the tree into three main  branches starting at the central site as highlighted by different colors  in Fig. \ref{figure1}. We denote with $\ket{l}_m$  the  normalized symmetric combination of local states defined on the nodes of the  $l^{th}$ generation in branch $m$, where $l=0$ denotes the central node of the Cayley tree. We  enumerate the
 three branches originating from the $l=0$ site with $m \in \{1, 2, 3\}$.\par
For example in Fig. \ref{figure1}
\begin{equation}
    \begin{split}
        \ket{2}_1 = \frac{1}{\sqrt{2}}(\ket{a}+\ket{b})\\
        \ket{2}_2 = \frac{1}{\sqrt{2}}(\ket{c}+\ket{d})\\
        \ket{2}_3 = \frac{1}{\sqrt{2}}(\ket{e}+\ket{f})
    \end{split}.
\end{equation}
	Note that there thus in total $3 M+1$ such symmetric basis states $\ket{l}_m$,
	which are orthogonal to each other. 
	
\begin{figure}[h]
    \centering
    \includegraphics[width=0.4\textwidth]{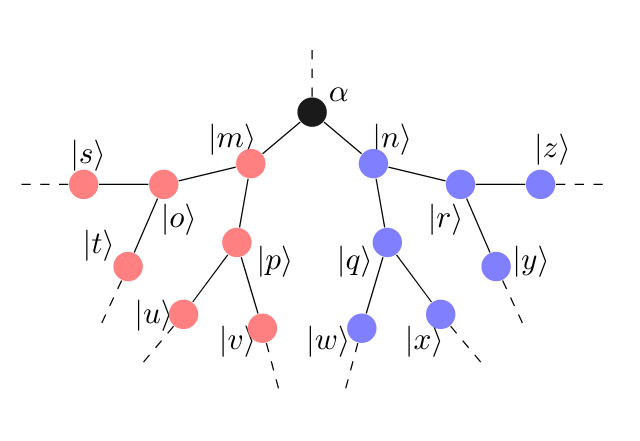}
    \caption{Two branches highlighted by blue and red,  arising from node $\alpha$. Forming  antisymmetric combinations of all sites on these two branches gives basis states.}
    \label{figure2}
\end{figure}

In order to get the remaining basis states, we 
include successively all antisymmetric superpositions of site states which branch from a node $\alpha$ in the $l^{th}$ generation of the Cayley tree to the right and left as shown in Fig. \ref{figure2}. The basis states are then taken to be the antisymmetric combination of states in the left and right branches evolving from  node $\alpha$, as highlighted by red and blue color in  Fig. \ref{figure2} (for each generation). Since there are $3\times 2^{l-1}$ states in each generation $l$, we enumerate these states with $\alpha \in \{1,2,...,3\times 2^{l-1}\}$.
Such a state starting in generation $l$ is thus  denoted as $\ket{l,r}_{\alpha}$ with 
$r\in \{1,..,M-l\}$. \\
For example, for the sites shown in
  Fig.\ref{figure2} (assuming that $\alpha$ is some node in $l^{th}$ generation of the cayley tree), the state labeled as $\ket{l,r=1}_{\alpha}$, $\ket{l,r=2}_{\alpha}$ and $\ket{l,r=3}_{\alpha}$ are given by
    \begin{align}
    \ket{l,1}_{\alpha} &=  \frac{1}{\sqrt 2} (\ket{m}-\ket{n})\\
      \ket{l,2}_{\alpha} &= \frac{1}{\sqrt{4}}(\ket{o}+\ket{p}-\ket{q}-\ket{r}) \nonumber \\
      \ket{l,3}_{\alpha} &= \frac{1}{\sqrt 8} (\ket{s} + \ket{t} + \ket{u} + \ket{v} - \ket{w}-\ket{x}-\ket{y}-\ket{z})\nonumber
    \end{align}
and so on for the further generations. Thus, for each node $\alpha$ in the $l^{th}$ generation we get $M-l$  such child states  $\ket{l,r}_{\alpha}$, with $r\in \{1,..,M-l\}$.  Since there are $3\times2^{l-1}$ nodes $\alpha$ in the $l^{th}$ generation, the total number of such states is,
\begin{equation}\label{Eq:gen_anti}
  \begin{split}
    \sum_{l = 1}^{M-1}{3 \times 2^{l-1}\left(M-l\right)} =3\left(2^M-1\right)-3M = N- 1- 3M.
  \end{split}
\end{equation}
Together with $3M+1$ symmetric basis states, we get in total $N$  basis states. Note that all antisymmetric states are orthogonal to one another and also to the symmetric states. One way of seeing this is as follows: a symmetric state always has equal amplitude on all nodes of a given branch, whereas an antisymmetric state always has equal number of nodes having positive and negative amplitudes (and can only have non-zero amplitude on nodes of one and the same branch). Thus, this completes the $N$ basis states forming an orthonormal basis for the Hilbert Space of the tight binding model on the Cayley tree with $N$ sites. Using this basis simplifies the solution of the eigenvalue equation Eq.  (\ref{Eq:beast2}), since it can be arranged in blocks, as we will see in the following section.
\subsection{Block Recurrence Relations}
Any eigenstate $\ket{\Psi}$ can now be written as a superposition of the basis states
\begin{equation}
\label{superposition}
\begin{split}
\ket{\Psi} &= \psi_0\ket{0}+\sum_{l=1}^{M}\sum_{m=1}^{3}\psi_{l,m}\ket{l}_m \\
&+ \sum_{l=1}^{M-1}\sum_{r=1}^{M-l}\sum_{\alpha\in \mathbb{G}_l}\phi^{\alpha}_{l,r}\ket{l,r}_{\alpha},
\end{split}
\end{equation}
where $\psi_{0}, \ \phi^{\alpha}_{l,r} \text{ and } \psi_{l,m}$ are complex amplitudes and $\mathbb{G}_l$ denotes the set of all $3\times2^{l-1}$ sites in the $l^{th}$ generation of the Cayley tree.

Insertion  of Eq. (\ref{superposition}) into the Eigenvalue equation Eq. (\ref{Eq:beast2})  results then in the  following recurrence relations
\begin{equation}\label{Eq:sym_recursions}
  \begin{split}
E\psi_0 &=  \psi_{1,1}+\psi_{1,2}+\psi_{1,3}, \\
E\psi_{l,m} &= \sqrt{2} (\psi_{l-1,m}+{\psi}_{l+1,m})\\
E{\psi}_{M,m} &= \sqrt{2} {\psi}_{M-1,m},
  \end{split}
\end{equation}
for $l=1,..., M-1$ and $m=1,2,3$, and
\begin{equation}\label{Eq:asym_recursions}
  \begin{split}
    E{\phi^{\alpha}_{l,1}} &=  \sqrt{2}{\phi^{\alpha}_{l,2}}, \\
E{\phi^{\alpha}_{l,r}} &= \sqrt{2} ({\phi^{\alpha}_{l,r-1}}+{\phi^{\alpha}_{l,r+1}})\\
E{\phi^{\alpha}_{l,M-l}} &= \sqrt{2}{\phi^{\alpha}_{l,M-l-1}}
  \end{split}
\end{equation}
for $l = 1,2...,M-1$ and $r=2,..., M-l-1$. \\

\subsection{Solutions of the recurrence relations}

Let us start with solutions which satisfy $\psi_0 = 0$, with which 
Eq.  (\ref{Eq:sym_recursions}) yields,
\begin{equation}\label{innerproduct}
  \psi_{1,1}+\psi_{1,2}+\psi_{1,3}=0,
\end{equation} 
and
\begin{equation}\label{OneRec}
 {\psi}_{l,m} = \left(\frac{E^2}{2}-1\right){\psi}_{l-2,m}-\frac{E}{\sqrt{2}}{\psi}_{l-3,m},
\end{equation}
for $l= 1,...,M.$
This gives,
\begin{equation}\label{Eq:psi_n}
  \begin{split}
 {\psi}_{l,m} &=\frac{-2^{\frac{1-3l}{2}} \left(\left(E-\sqrt{E^2-8}\right)^l-\left(\sqrt{E^2-8}+E\right)^l\right)}{\sqrt{E^2-8}}{\psi}_{1,m}\\
 &\equiv k_l{\psi}_{1,m},
 \end{split}
\end{equation}
for  $l= 1,...,M-1$. With the Ansatz 
for the Energy Eigenvalues $E = 2\sqrt{2}\cos\theta$, we get that
\begin{equation}\label{eq:k-n}
    k_l = \frac{1}{(2\sqrt{2})^{l-1}}\frac{\sin{l\theta}}{\sin\theta}.
\end{equation}
For now, $\psi_{1,m}$ can be freely choosen provided Eq. \ref{innerproduct} is satisfied. This will be fixed later on by requiring that the wavefunction be normalized. The equations in  Eq. (\ref{Eq:sym_recursions}) are closed by the open  boundary condition at the surface of the Cayley tree.  
 Open boundary condictions are implemented in the tight binding model 
by adding another generation of sites $M+1$, where the vanishing of the wave function
is imposed
\begin{equation}\label{Eq:bc}
  \begin{split}
 {\psi}_{M+1,m}&=0.  \\
  \end{split}
\end{equation}
We note that ${\psi_{1,m}}\neq{0}$ for at least one $m\in\{1,2,3\}$. Eq. (\ref{Eq:bc}) requires together with Eq. (\ref{Eq:psi_n}) the quantisation condition
\begin{equation}
\begin{split}
       k_{M+1} = 0
  ~\text{or} ~  \frac{\sin\{(M+1)\theta\}}{\sin{\theta}} = 0.  
\end{split}
\end{equation}
Thus, we get the following discrete solutions for $\theta$
\begin{equation}\label{Eq:quantization}
  \begin{split}
  \theta_i = \frac{\pi}{M+1}i,
    \end{split}
\end{equation}
with $i \in \{1,2,...,M\}$.\\
This gives the discrete energy eigenvalues
\begin{equation}\label{Eq:E-i}
    E_i = 2\sqrt{2}\cos\theta_i = 2\sqrt{2}\cos\left(\frac{\pi}{M+1}i\right).
\end{equation}
Next, we get $\psi_{1,m}$ by imposing 
the normalization condition
\begin{equation}
\sum_{l=1}^{M}\sum_{m=1}^{3}\abs{\psi_{l,m}}^2 = 1,
\end{equation}
which with Eq. (\ref{Eq:psi_n}) gives,
\begin{equation}\label{Eq:normalization}
  \begin{split}
    \sum_{l=1}^{N}\sum_{m=1}^{3}k_l^2\abs{\psi_{1,m}}^2=1 \\
    \text{or, } \sum_{m=1}^{3}\abs{\psi_{1,m}}^2= \frac{1}{\sum_lk_l^2}.
   \end{split}
\end{equation}
We can eliminate $\psi_{1,3}$ with Eq. (\ref{innerproduct}) to get,
\begin{equation}\label{Eq:ellipse}
\abs{\psi_{1,1}}^2+\abs{\psi_{1,2}}^2+\text{Re}(\psi_{1,1}\psi_{1,2}^*) = \frac{1}{2\sum_l{k_l^2}}.
\end{equation}
Defining $\psi_{1,1}:=r_1e^{i\nu_1}$ and $\psi_{1,2}:=r_1e^{i\nu_2}$ we get
\begin{equation}
  r_1^2+r_2^2+r_1r_2\cos{(\nu_1-\nu_2)}=\frac{1}{2\sum_l{k_l^2}},
\end{equation}
which, for  fixed energy $E$, gives a parameter family of ellipses for different $\Delta\nu:=\nu_1-\nu_2$.
For fixed energy $E$ we find the   two orthogonal solutions to Eq. \ref{Eq:ellipse}
\begin{equation}\label{Eq:psi-1-1}
\begin{split}
      \psi_{1,1}&=0,\ \psi_{1,2} = \frac{e^{i\mu}}{\sqrt{2}\sqrt{\sum_l{k_l^2}}} 
                                    \\
    \text{and }\psi_{1,1}&=\frac{e^{i\delta}}{\sqrt{2}\sqrt{\sum_l{k_l^2}}},\ \psi_{1,2} = 0,
\end{split}
\end{equation}
for arbitrary phases $\mu$ and $\delta$. 
Thus, all other  solutions of Eq. \ref{Eq:ellipse}  are linear combinations of these solutions. 
Next, using Eq. \ref{Eq:psi_n} and Eq. \ref{innerproduct}, we 
get all remaining complex amplitudes $\psi_{l,m}$.\par
Thus, for each possible energy eigenvalue $E_i$, given by Eq. (\ref{Eq:E-i}), we get two degenerate orthogonal eigenstates with the  following amplitudes on the basis vector components
\begin{equation}\label{eq:final-psi-n-m}
    \begin{split}
        \psi_{1,3} &= -\psi_{1,1}-\psi_{1,2}\\
        \text{and }\psi_{l,m} &= k_l\psi_{1,m}\\
        &= \frac{1}{(2\sqrt{2})^{l-1}}\frac{\sin{(l\theta_i)}}{\sin\theta_i} \psi_{1,m},
    \end{split}
\end{equation}
where the two possible choices of $\psi_{1,1}$ and $\psi_{1,2}$, as given by Eq. (\ref{Eq:psi-1-1}), give two orthogonal eigenstates with the  same energy $E_i$. Since there are $M$ possible values of $\theta_i$,  Eq. (\ref{Eq:quantization}) and each Eigenspace is two fold degenerate, the total number of states of this kind is $2M$.\\ \\

The Eigenstates given by  Eq. (\ref{eq:final-psi-n-m}) are orthogonal to  Mahan's symmetric
solutions, since the basis states in Mahan's analysis have equal weight in all three branches, and thus $\psi_{l,1}=\psi_{l,2}=\psi_{l,3}$ for all $l$. Since there are $M+1$ Mahan's solutions and they are orthogonal to the $2M$ solutions obtained above in Eq. (\ref{eq:final-psi-n-m}), we have obtained all the solutions we can get from  the 
subset of $3 M+1$ basis states obtained from  Eq. \ref{Eq:sym_recursions}.
We re-do the Mahan's analysis here. Solving Eq. (\ref{Eq:sym_recursions}) with the condition that $\psi_{l,1} = \psi_{l,2} = \psi_{l,3}$, we get that
\begin{equation}
\begin{split}
        \psi_{0,m} &= C\sin\gamma\\
    \psi_{l,m} &= C\sin(l\theta + \gamma) 
\end{split}
\end{equation}
where $E = 2\sqrt 2 \cos\theta$ and $\gamma$ is related to $\theta$ as:
\begin{align}\label{Eq:gamma-theta}
    \tan\gamma = 3 \tan\theta
\end{align}
$C$ is an overall normalization constant. The open boundary condition is imposed by adding another generation of sites and imposing that the wave function vanishes there, 
$\psi_{M+1,m} = 0$. This yields the 
quantization condition on $0<\theta \le \pi$ as 
\begin{equation}\label{eq:mahan-quantization}
\sin\{(M+1)\theta+\gamma\} = 0,
\end{equation}
or
\begin{align}\label{eq:mahan-quantization}
   \sin \{(M+1)\theta + \tan^{-1}(3 \tan\theta)\} = 0
\end{align}
This quantization condition, Eq. (\ref{eq:mahan-quantization}),
 has $M+1$ solutions with 
 $0< \theta \le \pi$, which can be found analytically   for given $M$.
\\ \\
Having solved the first set of recursion equations Eq. (\ref{Eq:sym_recursions}), we move on to solve the remaining Eq. (\ref{Eq:asym_recursions}). For given 
integer $l$ and $\alpha\in\mathbb{G}_l$, the second block  of equations 
in Eq. (\ref{Eq:asym_recursions}) resembles  the set of equations one obtains for the Eigenstates of a tight binding Hamliltonian on an  one-dimensional chain with $M-l$ sites. Thereby,  we can readily write its solutions as 
\begin{equation}\label{Eq:sol_chain}
  \begin{split}
  \phi^{\alpha}_{l,r} =& \sqrt{\frac{2}{M-l+1}}\sin{(r\chi)},
  \end{split}
\end{equation}
where  
 $r= 1,...,M-l$. The  energy eigenvalues are
\begin{equation}\label{Eq:E-beta}
E= 2\sqrt{2}\cos{\chi}.
\end{equation}
The possible values of $\chi$, for each choice of $l$ and $\alpha$, is 
obtained from the open boundary conditions $\phi_{M+1,r}^{\alpha}=0$,
yielding the quantisation condition
\begin{equation}\label{Eq:chi-i}
\begin{split}
    \chi_i = \frac{\pi}{M-l+1}i
    \text{~for~ } i = {1,2,...,M-l}.
\end{split}
\end{equation}
In particular, we obtain the {\it surface states} for $l = M-1$, where we get 
\begin{equation}\label{Eq:sol_chain_surface}
  \begin{split}
  \phi^{\alpha}_{M-1,r=1} = \sin{(\chi)}.
  \end{split}
\end{equation}
Eq. (\ref{Eq:chi-i}) dictates that 
$
    \chi = \frac{\pi}{2},$
and thus the Eigen energy of the surface states is 
$E=0$ with  eigenstates given by \begin{align}
    \ket{\Psi}_{\alpha}  = \ket{M-1,1}_{\alpha},
\end{align}
which is the antisymmetric combination of the two surface sites branching off from 
one of the $3\times2^{M-2}$ sites 
 $\alpha\in \mathbb{G}_{l=M-1}$.
Thus, we  showed that  antisymmetric combination of two  surface sites 
form the surface eigenstates of the Hamiltonian with zero energy,
$E=0$. The existence of such zero energy eigenstates can easily be verified directly from the form of the hamiltonian: taking two surface  states which are connected to the same site,
 one verifies that the structure of the Hamiltonian implies that 
 their antisymmetric state is a zero-energy eigenvector of $H$.

\begin{figure}[H]
    \centering
    \includegraphics[width=0.25\textwidth]{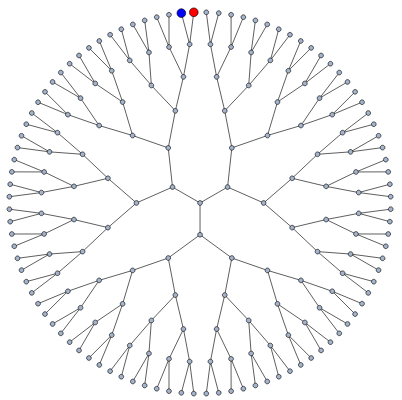}
    \caption{Visualisation of  a surface state, strongly localised 
with non-zero and equal intensity on two sites, only.  Red and blue indicate  opposite signs of the surface state amplitude.}
    \label{fig:surface}
\end{figure}

One of them is shown in Fig.  \ref{fig:surface}. For a Cayley tree with $M$-generations, there are  thus $N_{\rm Surface } = 3\times2^{M-2}$
such surface Eigenstates, which corresponds  for $M \gg 1$ to half of the total 
 states, $N_{\rm Surface }|_{M \gg 1 } \rightarrow N/4$.

 Having found all Eigenstates of the tight binding  Hamiltonian Eq.  (\ref{Eq:beast1}), 
let us  list them  in the following. 

\begin{figure}[H]
    \centering
    \includegraphics[width=0.25\textwidth]{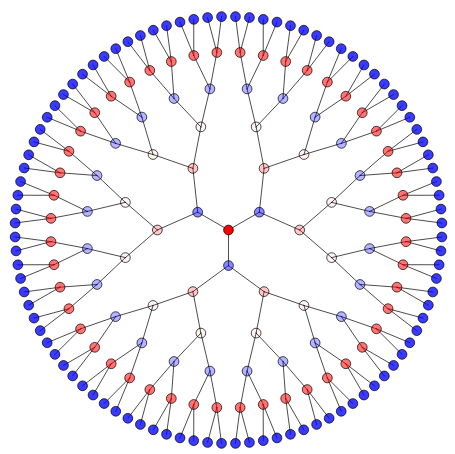}
    \caption{Visualisation of the amplitudes of a type I state for $M=6$ tree. The  darkness of the color is  proportional to the absolute value of the  Eigenstate  amplitude. Red denotes positive  and blue denotes negative amplitude. Note that the amplitude on sites of the same generation is equal.}
    \label{fig:mahan-state-pic}
\end{figure}

I. The $M+1$ symmetric states found by Mahan \cite{mahan}
with energy $E^I = 2\sqrt{2}\cos\theta_i,$ $i=0,1,...,M$
are given by 
\begin{eqnarray}
       && \ket{\Psi} = \psi_0\ket{0}+\sum_{l=1}^{M}\sum_{m=1}^{3}\psi_{l,m}\ket{l}_m\\
       &&   {\rm with~}  \psi_0 = C\sin{\gamma_i}
        {\rm ~,~}  \psi_{l,m} = C\sin({l\theta_i+\gamma_i}),
\end{eqnarray}
where the $M+1$ solutions  $0<\theta_i \le \pi$ are obtained from  the condition Eq. (\ref{eq:mahan-quantization}) and
the $\gamma_i$ is defined by Eq. (\ref{Eq:gamma-theta}).
These states are extended, have equal amplitudes in nodes of the same generation and have finite amplitude  $\psi_0$ at the centre of the tree.
 The type I state of lowest energy is obtained by setting the argument of $\sin $
 in  the condition Eq. (\ref{eq:mahan-quantization}) equal to $M \pi$, yielding 
 the energy to be $E^I_0 = 2 \sqrt{2} \cos \theta_0$ with $\theta_0$ being a solution of the equation 
  \begin{equation}  \label{t0}
  \theta_0 = \frac{M}{M+1}\pi - \frac{1}{M+1} \tan^{-1} (3\tan\theta)).
  \end{equation}
   We find that $\frac{M}{M+1}\pi < \theta_0 < \pi$, for all $M>1$,
   since $\tan^{-1} (3\tan\theta) <0$ for $\pi/2 < \theta_0 < \pi$,
   and the solution  $\theta_0 = \pi$ corresponds to the trivial solution 
   with wavefunction $\psi =0$ which is  not an Eigenstate.
  Thus, we find that 
   the Eigen energy  $E^I_0 = 2 \sqrt{2} \cos \theta_0$ is the smallest 
    Eigen energy of all states, including the 
    other  type II and type III states and 
   we can conclude that the ground state is for any $M$
   the  extended symmetric type I state with energy 
   \begin{equation} \label{groundstate}
   E^I_0 = 2 \sqrt{2} \cos \theta_0,
   \end{equation} 
   where $\theta_0$ is the nontrivial solution of Eq. (\ref{t0}).
   The ground state for a $M=6$ Cayley tree is shown in Fig. \ref{fig:mahan-state-pic}.
  
\begin{figure}[H]
    \centering
    \includegraphics[width=0.25\textwidth]{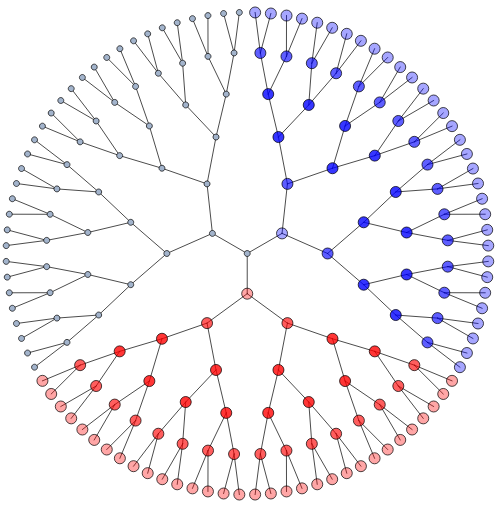}
    \caption{Visualisation of the amplitudes  of a Type II state with $\psi_{1,1} = 0$. The  darkness of the color is  proportional to the absolute value of the  Eigenstate  amplitude.
   Red denotes positive  and blue denotes negative amplitude. Note that the central site has zero amplitude and
    the amplitude on sites of the same generation on each branch is equal. }
    \label{fig:typeII}
\end{figure}

\begin{figure}[H]
    \centering
    \includegraphics[width=0.25\textwidth]{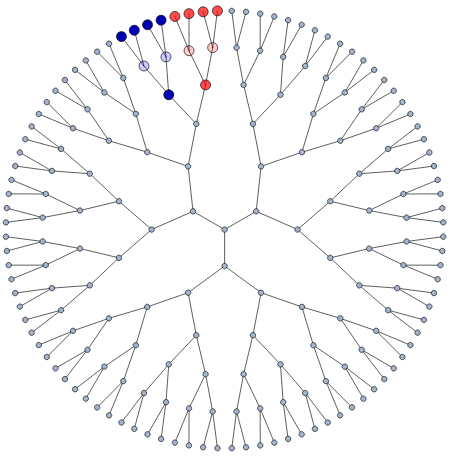}
    \caption{Visualisation of the amplitudes  of a Type III state for $l=3$ and $M=6$. The  darkness of the color is  proportional to the absolute value of the  Eigenstate  amplitude. Red denotes positive  and blue denotes negative amplitude.  Note that the amplitude on sites of the same generation is equal.}
    \label{fig:typeIII}
\end{figure}

II. Next, there are $2 M$ states with energies
$E^{II}_i = 2\sqrt{2}\cos\theta_i = 2\sqrt{2}\cos{\left(\frac{\pi}{M+1}i\right)}$,
where $i\in\{1,2..,M\}$,
given by 
 \begin{gather}
            \ket{\Psi_i} = \sum_{l=1}^{M}\sum_{m=1}^{3}\psi_{l,m}\ket{l}_m 
          \end{gather}  
           { with }
           $\psi_{1,3} = -\psi_{1,1}-\psi_{1,2}$,
    and $    \psi_{l,m}=\frac{1}{(2\sqrt{2})^{l-1}}\frac{\sin{l\theta_i}}{\sin\theta_i} \psi_{1,m}\\
$
where  each energy $E^{II}_i$ is two fold degenerate with  two orthogonal states given by $(\psi_{1,1},\psi_{1,2})$ from Eq. (\ref{Eq:psi-1-1}).  These $2M$ states are extended throughout the Cayley tree, except that they  have zero amplitude at the centre of the tree.
An example is shown in Fig. \ref{fig:typeII}.
We note that the type II states with lowest energy are the two states with $i =M$ yielding 
 the energy $E^{II}_M = 2\sqrt{2}\cos{\left(\frac{\pi}{M+1}M\right)}$. 
III. The remaining states with energeis  $E^{III}_i = 2\sqrt{2}\cos\chi_i = 2\sqrt{2}\cos{\left(\frac{\pi}{M-l+1}i\right)}$ with 
        $i \in\{1,2,...,M-l\}$ and $l\in\{1,2,...,M-1\}$ are given by
\begin{equation}
\ket{\Psi}_{l,\alpha} = \sum_{r=1}^{M-l}\phi^{\alpha}_{l,r}\ket{l,r}_{\alpha},
\end{equation}
with
\begin{equation}
        \phi^{\alpha}_{l,r} = \sqrt{\frac{2}{M-l+1}}\sin{(r\chi_i)},
\end{equation}
where  $l\in\{1,2,...,M-1\}$ and $\alpha\in\mathbb{G}_l$ and 
 $r= 1,...,M-l$. For fixed $l$ and $\alpha$, there are thus $M-l$ possible values of $E_i = 2\sqrt{2}\cos\chi_i$ with $\chi_i = \frac{\pi}{M-l+1}i$
    for $ i = {1,2,...,M-l}$. Since there are $M$ possible values of $l$ and there are $3\times 2^{l-1}$ nodes $\alpha$ in $\mathbb{G}_l$, by similar computation as in Eq. (\ref{Eq:gen_anti}), we get that there are $N-(3M+1)$ states of this type. These states are localized to branches of the Caley tree and for increasing $l$ they get more and more localized, with finite amplitude only on $2^{M-l}$ sites. An example 
     of  a Type III state for $l=3$ and $M=6$ is shown in Fig. \ref{fig:typeIII}.
 The  $3 \times 2^{M-2}$ surface states are  antisymmetric combinations of two sites 
 with Eigenergy $E=0$, exactly. 
 The  type III state with lowest energy is obatined for $ l= 1$ and $i= M-1$ yielding the energy
 \begin{equation} \label{localised}
 E^{III}_{l=1,i=M-1} =  2\sqrt{2}\cos{\left(\frac{\pi}{M}(M-1)\right)},
 \end{equation}
  which is 
  always larger than energy $E^I_0$ of   the extended ground state    and
  energy $E^{II}_M$ of the first excited, extended states. 
We note that these results are in complete agreement with the numerical 
results of Ref.  \cite{Yorikawa2018}, where numerical results for $M=4$ have been presented. 
We also note that one can distinguish states 
with energy $E = 2 \sqrt{2} \cos \theta$ with $\theta$ 
being  a rational fraction of $\pi$, and others where  $\theta$ is  
  an irrational fraction of $\pi$. But we find that 
this classification is not sufficient to distinguish the different nature of the Eigenstates 
as we have classified and identified them:  The  extended type II states and the localised type III states
 have both an energy $E = 2 \sqrt{2} \cos \theta$ where $\theta$  is  a rational fraction of $\pi$. The type I states can have $\theta$ being an  irrational fraction of $\pi$,
 but may also be  a rational fraction of $\pi$. Thus, this irriationality is not a necessary condition for a type I state.

\section{Density of  States}
Mahan had calculated in Ref. \onlinecite{mahan} the local density of states at the central site  
  $\rho_{00} (E)$. Since 
   only the $M+1$ shell symmetric states have a finite amplitude $\psi_0$ at that site one finds\cite{mahan}
   \begin{equation}
        \rho_{00} (E) = \sum_{\theta_n}  |\psi_0 (\theta_n)|^2\delta(E - \epsilon_{n}),
\end{equation}
where $\epsilon_{n}= 2 \sqrt{2} \cos{\theta_n}$.
 The summation can be approximated by an integral over $\theta$ in the limit of a Bethe lattice, $M\rightarrow \infty$,
 where one finds\cite{mahan}
  \begin{equation}
        \rho_{00} (E) = \int_0^{\pi} d \theta \frac{2}{\pi} {\sin}^2(\beta(\theta)) \delta(E - \epsilon_{n}) = \frac{3}{2 \pi} \frac{\sqrt{8-E^2}}{9-E^2}.
\end{equation}
Having obtained all the eigenstates and energies of the Schroedinger equation on the Cayley tree, we can now proceed to calculate the total density of states, given by 
\begin{equation}
        \rho (E) = \sum_n \delta(E - \epsilon_{n})
\end{equation}
where the sum is over all Eigen energies $\epsilon_n$.
It is convenient to write it as a sum of contributions from the
threee different different kind of states $I,II,III$, we have derived above,
\begin{equation}
        \rho (E) = \rho_{I} (E) + \rho_{II} (E) + \rho_{III}(E)
\end{equation}
where $\rho_{I}$ denotes the contribution due to the  $M+1$ symmetric Mahan states, $\rho_{II}$  the  contribution due to the $2M$ states which have same amplitude in each of the three branches  and $\rho_{III}$ the contribution due to the $N-(3M+1)$  states
which are localised in different branches of the Cayley tree. 
In the large $M$ limit, we get
\begin{equation}
        \rho_I(E) = \frac{M+1}{\pi\sqrt{8-E^2}},
\end{equation}
see the Appendix for the derivation. 
Similarly, we find
\begin{equation}
    \begin{split}
    \rho_{II}(E) = \frac{2(M+1)}{\pi\sqrt{8-E^2}},
    \end{split}
\end{equation}
and
\begin{equation}
    \begin{split}
    \rho_{III}(E) = \sum_{l=1}^{M-1}\mathbb{D}_l\sum_{i=1}^{M-l}\ \delta\left\{E-2\sqrt{2}\cos\left(\frac{\pi}{M-l+1}i\right)\right\},
    \end{split}
\end{equation}
where \(\mathbb{D}_l = \sum_{\alpha\in\mathbb{G}_l}\) = \(3\times 2^{l-1}\) is the number of sites in the \(l^{th}\) generation. We observe that the 
degeneracy of the states increases with $M$ 
as $\sim 2^M$ for the $III-$type  Eigenstates which are 
localized near the surface. 
Thus, in the $M\rightarrow \infty$ limit, those states are highly degenerate.

\begin{figure}[H]
    \centering
    \includegraphics[width=0.35\textwidth]{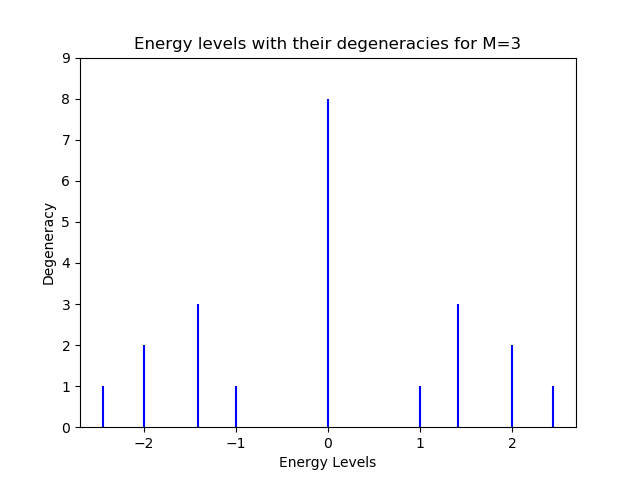}
    \caption{Histogram shows the degeneracy of energy levels computed numerically for $M=3$.}
    \label{fig:histo}
\end{figure}

As an example we show the results for  a 
 numerical computation of the density of states  for  a Cayley tree with $M=3$ generations.
 We  see the distribution of energy eigenvalues and their degeneracies in  Fig. \ref{fig:histo}, in particular  that the states at  zero energy $E=0$ are
 highly degenerate. The histogram is symmetric about $E = 0$, with same number of states with energy $-E$ as with energy $E$ as is clear from 
  the analytical solution Eqs. \ref{Eq:quantization}, \ref{eq:mahan-quantization}, \ref{Eq:chi-i} for $M=3$.



\section{Conclusion}
 The complete set of Eigenstates and Eigenvalues of the nearest neighbour tight binding model on a Cayley tree with branching number $b=2$ and $M$ branching generations with open boundary conditions has been  derived. 
  Besides the $M+1$ shell symmetric Eigenstates derived already by Mahan in Ref. \onlinecite{mahan},
   we find $2 M$ Eigenstates which have zero amplitude at central site but  are otherwise extended throughout the Cayley tree. The remaining $N-(3 M+1)$ states are found to be strongly localised states with finite amplitudes on only a subset of sites. In particular, there are  $3 \times 2^{M-2}$
    states which are  each antisymmetric combinations of two sites at the surface of the Cayley tree. Thus for $N\gg1$, there 
     are $N_{\rm surface} \rightarrow N/4$ surface states. We find that the ground state of the Cayley tree is always an extended symmetric state with 
     ground state energy given by Eq. (\ref{groundstate}). The two first excited states are also 
      extended states. The localised states are spread over the spectrum.
       The most strongly localised states have energy $E=0$ and are localised on a pair of neighboured surface states, being an antisymmetric superposition of them. 
         The localised states with   the largest localisation length of order $M$ sites have  the
         lowest  energy of all localised states, given by Eq. (\ref{localised}). 
         
 These results are used to derive  the  total density of states as function of energy $E$. Having all Eigenstates and
 Eigenfunctions one can
 now derive the local density of states not only at the central site as in Ref. \onlinecite{mahan} but at any site and any one particle observable on the Cayley tree, as well as one particle response functions. Using a similar strategy, choosing a convenient basis, we can derive the Eigenstates for any
  branching number $b>2,$ a task we leave for a future publication.  
  
  After submission of this manuscript, we were made aware of Ref. \cite{Yorikawa2018} who studied the same problem, and derived the Eigenstates and Eigenvalues of the Cayley tree numerically, 
  with results which are in agreement with our analytical solutions, as presented here. 
\acknowledgments
S.K. gratefully acknowledges support from DFG KE-807/22-1.
 
\appendix*
\section{}

{\it Derivation of $\rho_I(E).$}
With  the dispersion relation 
 $$E_n = 2\sqrt{2}\cos{\theta_n},$$ and  
the  quantization condition 
\begin{align*}
    \theta_n +\frac{\gamma(\theta_n)}{M+1} &= \frac{\pi n}{M+1},
\end{align*}
 we obtain, using  $\displaystyle{\sin^2{\gamma} = \frac{9\sin^2{\theta}}{9-8\cos^2{\theta}}}$, the phase differences
\begin{equation}\label{Eq:theta-diff}
    \theta_{n+1}-\theta_n = \frac{\pi}{M+1}-\frac{\gamma(\theta_{n+1})-\gamma(
\theta_n)}{M+1}.
\end{equation}
Thus, for $M\rightarrow \infty$ we 
can approximate the ratio of  differences  of $\gamma$ and $\theta$ by the first  derivative of $\gamma$ with respect to $\theta$,
$$\frac{\gamma(\theta_{n+1})-\gamma(\theta_n)}{\theta_{n+1}-\theta_n} \approx \frac{\partial \gamma(\theta)}{\partial\theta}.$$
Now, writing $x = \sin \theta $ and  $y = \sin \gamma$ we obtain 
$ y^2 = 9x^2/(1+8x^2)$  which implies 
$$ \frac{\partial y}{\partial x} = \frac{3}{(1+8x^2)^{3/2}}.$$
Thus,  $$\frac{\partial y}{\partial x} = \frac{\partial \sin{\gamma}}{\partial\sin{\theta}} = \frac{ \cos{\gamma}}{\cos{\theta}}\ \frac{\partial \gamma}{\partial \theta}, $$
 and $$ \frac{\partial \gamma}{\partial \theta} = \frac{\cos{\theta}}{\cos{\gamma}}\ \frac{\partial y}{\partial x} = \frac{\cos{\theta}}{\cos{\gamma}}\ \frac{3}{(1+8\sin^2{\theta})^{3/2}}. $$
Thereby,  Eq. \ref{Eq:theta-diff} becomes for $M \gg 1$,
\begin{align*}
\theta_{n+1}-\theta_{n} &= \frac{\pi}{M+1}-\frac{\gamma(\theta_{n+1})-\gamma(\theta_{n})}{\theta_{n+1}-\theta_n} \  \frac{\theta_{n+1}-\theta_n}{M+1}\\
                        &\approx \frac{\pi}{M+1}\left\{1-\frac{3}{M+1}\frac{\cos{\theta_n}}{\cos{\gamma_n}}\frac{1}{(1+8\sin^2{\theta_n})^{3/2}} \right\}\\
                        &+\text{o}\left(\frac{1}{(M+1)^3}\right).
\end{align*}
For \(M\gg1\), we thus find 
with $\theta_{n+1}-\theta_{n} \to \frac{\pi}{M+1}$  that 
 the contribution of the $I-$ states to the  density of states is given by 
\begin{align*}
\rho_{I}(E) 
	     &= \frac{M+1}{\pi}\int_{0}^{\pi}d\theta\ \delta(E-2\sqrt{2}\cos\theta)\\
	     &= \frac{M+1}{\pi}\int_{\varepsilon(0)}^{\varepsilon(\pi)}d\varepsilon\ \frac{1}{\left|\frac{d\varepsilon}{d\theta}\right|}\delta(E-\varepsilon).
\end{align*}
With $$\frac{d\varepsilon}{d\theta} = 2\sqrt{2}\sin\theta = 2\sqrt{2}\sqrt{1-\cos^2\theta} = 2\sqrt{2}\sqrt{1-\frac{\varepsilon^2}{8}},$$  we finally find
\begin{equation}
\rho_{I}(E) = \frac{M+1}{\pi\sqrt{8-E^2}}.
\end{equation}

{\it Derivation  of $\rho_{II}$.}
With the dispersion relation 
$E = 2\sqrt{2}\cos(\theta_i)$ and the quantisation condition $ \theta_i = \frac{\pi}{M+1}i.$
 we find for $M \gg1 $ in the continuum limit and noting that each energy level 
  is 2-fold degenerate,
  \begin{equation}
  \rho_{II} (E) = \frac{2(M+1)}{\pi\sqrt{8-E^2}},
  \end{equation}
twice the contribution  as for the shell symmetric $I-$ states.

{\it Derivation  of $\rho_{III}$.}
With the  dispersion relation 
$E_{i,l} = 2\sqrt{2}\cos\left({\frac{\pi}{M-l+1}}i\right)$
for $i = 1,2,....,M-l$ and $l = 1,2,...,M-1$ we find
\begin{align*}
\rho_{III}(E) &= \sum_{l=1}^{M-1}\sum_{\alpha\in \mathbb{G}_l}\sum_{i=1}^{M-l}\delta\left\{E-2\sqrt{2}\cos\left(\frac{\pi}{M-l+1}i\right)\right\}\\
	      &= \sum_{l=1}^{M-1}\mathbb{D}_l\sum_{i=1}^{M-l}\ \delta\left\{E-2\sqrt{2}\cos\left(\frac{\pi}{M-l+1}i\right)\right\}
\end{align*}
where \(\mathbb{D}_l = \sum_{\alpha\in\mathbb{G}_l}\) is the
number of sites in the \(l^{th}\) generation,  \( \mathbb{D}_l =3\times 2^{l-1}\). \\

\newpage


\end{document}